\input harvmac
\input epsf
\input diagrams
\noblackbox
\newcount\figno
\figno=0
\def\fig#1#2#3{
\par\begingroup\parindent=0pt\leftskip=1cm\rightskip=1cm\parindent=0pt
\baselineskip=11pt
\global\advance\figno by 1
\midinsert
\epsfxsize=#3
\centerline{\epsfbox{#2}}
\vskip 12pt
\centerline{{\bf Figure \the\figno} #1}\par
\endinsert\endgroup\par}
\def\figlabel#1{\xdef#1{\the\figno}}
\def\pano{\par\noindent}
\def\smno{\smallskip\noindent}
\def\meno{\medskip\noindent}
\def\bigno{\bigskip\noindent}
\font\cmss=cmss10
\font\cmsss=cmss10 at 7pt
\def\rlx{\relax\leavevmode}
\def\inbar{\vrule height1.5ex width.4pt depth0pt}
\def\IC{\relax\,\hbox{$\inbar\kern-.3em{\rm C}$}}
\def\IR{\relax{\rm I\kern-.18em R}}
\def\IN{\relax{\rm I\kern-.18em N}}
\def\IP{\relax{\rm I\kern-.18em P}}
\def\ZZ{\rlx\leavevmode\ifmmode\mathchoice{\hbox{\cmss Z\kern-.4em Z}}
 {\hbox{\cmss Z\kern-.4em Z}}{\lower.9pt\hbox{\cmsss Z\kern-.36em Z}}
 {\lower1.2pt\hbox{\cmsss Z\kern-.36em Z}}\else{\cmss Z\kern-.4em Z}\fi}
\def\narrowplus{\kern -.04truein + \kern -.03truein}
\def\narrowminus{- \kern -.04truein}
\def\narrowminussub{\kern -.02truein - \kern -.01truein}

\def\o#1{\overline{#1}}

\def\th#1#2{\vartheta\bigl[{  #1 \atop #2} \bigr] }



\lref\rsagbi{A. Sagnotti, M. Bianchi,
{\it On the Systematics of Open String Theories},
Phys.\ Lett.\ {\bf B247} (1990) 517.}

\lref\rseiwit{N. Seiberg and  E. Witten, {\it String Theory and 
Noncommutative Geometry}, JHEP {\bf 9909} (1999) 032, hep-th/9908142.}

\lref\rfgrs{J. Fr\"ohlich, O. Grandjean, A. Recknagel and V. Schomerus, 
{\it Fundamental Strings in Dp-Dq Systems}, hep-th/9912079.}

\lref\rbdl{M. Berkooz, M. R. Douglas and R. G. Leigh, 
{\it Branes intersecting at Angles}, Nucl. Phys. {\bf B480} (1996) 265, 
hep-th/9606139.}

\lref\rfgr{J. Fr\"ohlich, O. Grandjean and A. Recknagel, 
{\it Supersymmetric Quantum Theory and Noncommutative Geometry}, 
Commun. Math. Phys. {\bf 203} (1999) 119, math-ph/9807006.}

\lref\rjabbari{M. M. Sheikh-Jabbari, 
{\it More on mixed Boundary Conditions and D-branes Bound States}, Phys. Lett. 
{\bf B425} (1998) 48, hep-th/9712199.}

\lref\rbrunner{I. Brunner, A. Rajaraman and M. Rozali, 
{\it D-branes on Asymmetric Orbifolds}, Nucl. Phys. {\bf B558} (1999) 205, 
hep-th/9905024.} 

\lref\rkaku{ Z. Kakushadze, G. Shiu and  S.-H. H. Tye, {\it
   Type IIB Orientifolds, F-theory, Type I Strings on Orbifolds and 
    Type I - Heterotic Duality}, Nucl. Phys. {\bf B533} (1998) 25,
    hep-th/9804092.}

\lref\rdine{M. Dine and E. Silverstein, {\it New M-theory Backgrounds with 
Frozen Moduli}, hep-th/9712166.}

\lref\rbg{R. Blumenhagen and L. G\"orlich, {\it Orientifolds of 
Non-Supersymmetric Asymmetric Orbifolds}, Nucl.Phys. {\bf B551} (1999) 601,
hep-th/9812158.}

\lref\rbgka{R. Blumenhagen, L. G\"orlich and B. K\"ors, 
{\it Supersymmetric Orientifolds 
in 6D with D-Branes at Angles}, hep-th/9908130; {\it 
A New Class of Supersymmetric Orientifolds with D-Branes at Angles}, 
hep-th/0002146.}

\lref\rbgkb{R. Blumenhagen, L. G\"orlich and B. K\"ors, {\it Supersymmetric 4D
 Orientifolds of Type IIA with D6-branes at Angles}, 
JHEP {\bf 0001} (2000) 040, hep-th/9912204.}

\lref\rba{R. Blumenhagen and C. Angelantonj, {\it Discrete Deformations in 
Type I Vacua}, Phys. Lett. {\bf 473} (2000) 86, hep-th/9911190.}

\lref\rprad{G. Pradisi, {\it Type I Vacua from Diagonal $Z_3$-Orbifolds},
   hep-th/9912218.}

\lref\rbimopra{M. Bianchi, J.F. Morales and  G. Pradisi, {\it Discrete torsion 
in non-geometric orbifolds and their open-string descendants}, 
hep-th/9910228.}

\lref\rdlp{J. Dai, R.G. Leigh and J. Polchinski, {\it New Connections
Between String Theories},  Mod.Phys.Lett. {\bf A4} (1989) 2073.}

\lref\rnappi{C.G. Callan, C. Lovelace, C.R. Nappi and S.A. Yost, {\it
String Loop Corrections To Beta Functions}, Nucl. Phys. {\bf B288}
(1987) 525;
A. Abouelsaood, C.G. Callan, C.R. Nappi and S.A. Yost, 
{\it Open Strings in Background Gauge Fields}, Nucl. Phys. {\bf B280}
(1987) 599.}

\lref\rconnes{A. Connes, M.R. Douglas and  A. Schwarz, {\it 
Noncommutative Geometry and Matrix Theory: Compactification on Tori},
JHEP {\bf 9802} (1998) 003, hep-th/9711162;
M.R. Douglas and  C. Hull, {\it
D-branes and the Noncommutative Torus}, JHEP {\bf 9802} (1998) 008,
hep-th/9711165.}

\lref\rkrogh{Y.-K.E. Cheung and M. Krogh, {\it Noncommutative Geometry from 
D0-branes in a Background B-field}, Nucl.Phys. {\bf B528} (1998) 185,
hep-th/9803031.}

\lref\rschomi{V. Schomerus, {\it D-branes and Deformation Quantization},
JHEP {\bf 9906} (1999) 030, hep-th/9903205.}

\lref\rfroh{J. Fr\"ohlich, O. Grandjean, A. Recknagel, {\it
Supersymmetric quantum theory and non-commutative geometry},
Commun.Math.Phys. 203 (1999) 119, math-ph/9807006.}

\lref\raas{F. Ardalan, H. Arfaei and  M. M. Sheikh-Jabbari, {\it 
Noncommutative Geometry From Strings and Branes}, 
JHEP {\bf 9902} (1999) 016, hep-th/9810072.}

\lref\rchu{C.-S. Chu, {\it Noncommutative Open String: Neutral and Charged},
hep-th/0001144.}

\lref\rhho{C.-S. Chu and  P.-M. Ho, 
{\it Noncommutative Open String and D-brane},  
Nucl.Phys. {\bf B550} (1999) 151, hep-th/9812219;
{\it Constrained Quantization of Open 
String in Background B Field and Noncommutative D-brane}, hep-th/9906192.}

\lref\rchen{B. Chen, H. Itoyama, T. Matsuo and K. Murakami, {\it
p-p' System with B-field, Branes at Angles and Noncommutative Geometry}, 
hep-th/9910263.}

\lref\rlee{T.  Lee, {\it Canonical Quantization of Open String and
Noncommutative Geometry},  hep-th/9911140.}

\lref\rbound{H. Arfaei and D. Kamani, {\it Branes with Background Fields 
in Boundary State Formalism}, Phys.Lett. {\bf B452} (1999) 54, 
hep-th/9909167.}

\lref\raasb{F. Ardalan, H. Arfaei, M. M. Sheikh-Jabbari, {\it
Dirac Quantization of Open Strings and Noncommutativity in Branes},
hep-th/9906161.}

\lref\rvafa{K.S. Narain, M.H. Sarmadi and C. Vafa, {\it
   Asymmetric Orbifolds}, Nucl.Phys. {\bf B288} (1987) 551.}

\lref\rangles{M. Berkooz, M. R. Douglas and  R.G. Leigh, {\it Branes 
Intersecting at Angles}, Nucl.Phys. {\bf B480} (1996) 265, hep-th/9606139;
H. Arfaei and M. M. Sheikh Jabbari, {\it Different D-brane Interactions},
Phys.Lett. {\bf B394} (1997) 288, hep-th/9608167;
J. C. Breckenridge, G. Michaud and R. C. Myers, {\it New angles on D-branes},
Phys.Rev. {\bf D56} (1997) 5172, hep-th/9703041;
M. M. Sheikh Jabbari, {\it Classification of Different Branes at Angles},
Phys.Lett. {\bf B420} (1998) 279, hep-th/9710121.}

\Title{\vbox{\hbox{hep--th/0003024}
 \hbox{HUB--EP--00/15}}}
{\vbox{\centerline{Asymmetric Orbifolds, Noncommutative Geometry}
\bigskip\centerline{and Type I String Vacua}
}}
\centerline{Ralph Blumenhagen, Lars G\"orlich, 
Boris K\"ors and Dieter L\"ust}
\bigskip
\centerline{\it 
Humboldt-Universit\"at zu Berlin, Institut f\"ur 
Physik,}
\centerline{\it \   Invalidenstrasse 110, 10115 Berlin, Germany }
\smallskip
\bigskip
\centerline{\bf Abstract}
\noindent
We investigate the D-brane contents of asymmetric orbifolds. 
Using T-duality we find that the consistent description of open
strings in asymmetric orbifolds requires to turn
on background gauge fields on the D-branes. 
We derive the corresponding noncommutative geometry arising
on such D-branes with mixed Neumann-Dirichlet boundary conditions 
directly by applying
an asymmetric rotation to open strings with pure Dirichlet or Neumann 
boundary conditions. As a concrete application of our results
we construct asymmetric type I vacua requiring open strings
with mixed boundary conditions for tadpole cancellation. 
\footnote{}
{\pano
e--mail:\ blumenha, goerlich, koers, luest@physik.hu-berlin.de
}
\Date{03/2000}

\newsec{Introduction}

As is known since the work of Connes, Douglas and Schwarz \rconnes,
matrix theory compactifications on tori with background
three-form flux lead to noncommutative geometry.  
Starting with the early work \rnappi\ one has  subsequently realized
that open strings moving in backgrounds with non-zero
two-form flux or non-zero gauge fields  have mixed boundary conditions
leading  to a noncommutative geometry on the boundary of the
string world-sheet 
\refs{\rkrogh\rfgr\raas\rhho\rschomi\raasb\rseiwit\rbound\rchen\rlee\rfgrs-\rchu}.
As pointed out in \rseiwit,
also the effective theory on the D-branes becomes a noncommutative
Yang-Mills theory. 

We know from the discovery of D-branes, that Dirichlet
branes made their first appearance by studying the realization of T-duality
on a circle in the open string sector \rdlp. For instance, 
starting with a D9 brane, the
application of T-duality leads to a D8-brane where the ninth direction
changes from a Neumann boundary condition to a Dirichlet boundary 
condition. Thus, one may pose the question how D-branes with 
mixed Neumann-Dirichlet
boundary conditions fit into this picture. Does there exist
a transformation relating pure Dirichlet or Neumann boundary conditions to
mixed Neumann-Dirichlet
boundary conditions? 

At first sight unrelated, there exists the so far unresolved problem of 
what the D-brane content of asymmetric orbifolds is. The simplest asymmetric
orbifold is defined by 
modding out by T-duality itself, which is indeed a symmetry as long
as one chooses the circle at the self-dual radius. Thus,
as was argued in \rdine\ and applied to type I compactifications
in \rbg, in this special case D9- and D5-branes are identified
under the asymmetric orbifold action. However, the general T-duality
group for compactifications on higher dimensional tori contains
more general asymmetric operations. For instance, the 
root lattice of $SU(3)$  allows  an asymmetric $\widehat{\ZZ}_3$ action.
(A left-right asymmetric $\ZZ_N$ symmetry is denoted by $\widehat{\ZZ}_N$.)
The closed string sector can very well live with such non-geometric
symmetries \rvafa\ but what about the open string sector? Since all type II
string theories contain open strings in the non-perturbative D-brane sector,
in order for asymmetric orbifolds to be non-perturbatively consistent,
one has to find a realization of such non-geometric symmetries
in the open string sector, as well. Thus, the question arises what
the image of a D9-brane under an asymmetric $\widehat{\ZZ}_N$ action
is. 

The third motivation for the investigation performed in this paper is
due to recently introduced orientifolds with D-branes at angles 
\refs{\rbdl\rbgka\rba\rbgkb-\rprad}. 
We investigated orientifold models for which the world-sheet parity 
transformation,
$\Omega$, is combined with a complex conjugation, ${\cal R}$,
of the compact coordinates. After dividing by a further left-right symmetric
$\ZZ_N$
space-time symmetry the cancellation of  tadpoles
required the  introduction of  so-called  twisted
open string sectors. These sectors were realized by open strings stretching
between 
D-branes intersecting at non-trivial angles. As was pointed out
in \rbgka, these models are related to ordinary $\Omega$ orientifolds
by T-duality. However, under this T-duality the former left-right
symmetric $\ZZ_N$ action is turned into an asymmetric $\widehat{\ZZ}_N$
action in the dual model. Thus, we are led to the problem of
describing asymmetric orientifolds in a D-brane language. 
Note, that using pure conformal field theory methods asymmetric
orientifolds were discussed recently in \rbimopra.

In this paper,  we study the three 
conceptually important problems mentioned
above, for simplicity, in the case of  compactifications on direct 
products of two-dimensional tori.
It turns out that all three problems are deeply related. 
The upshot is that asymmetric rotations turn Neumann boundary conditions
into mixed Neumann-Dirichlet
boundary conditions. This statement is  the solution to the first
problem and allows us to rederive the noncommutative geometry
arising on D-branes with background gauge fields simply by
applying asymmetric rotations to ordinary D-branes.
The solution to the second problem is that asymmetric orbifolds
necessarily contain open strings with mixed boundary conditions. 
In other words: D-branes manage to incorporate asymmetric symmetries 
by turning on background gauge fluxes, which renders their world-volume 
geometry noncommutative. Gauging the asymmetric symmetry 
can then lead to an identification of commutative and 
noncommutative geometries. In this sense 
asymmetric type II orbifolds are deeply related to noncommutative geometry. 
Apparently, the same holds for asymmetric
orientifolds, orbifolds of type I. Via T-duality the whole plethora of 
$\Omega {\cal R}$ orientifold models of \refs{\rbgka\rba-\rbgkb} 
is translated into a set of 
asymmetric orientifolds with D-branes of different commutative and 
noncommutative types in the background. We will further  
present a D-brane interpretation
of some of the non-geometric models studied in \rbimopra\ and some
generalizations thereof. 

In section 2 we describe a special class  of asymmetric 
orbifolds on $T^2$. Employing T-duality we first determine
the tori allowing an asymmetric $\widehat{\ZZ}_N$ action, where we 
discuss the   $\widehat{\ZZ}_3$ example in some detail. Afterwards
we study D-branes in such models and also determine the zero-mode
spectrum for some special values of the background gauge 
flux.
In section 3 we apply asymmetric rotations to give an alternative
derivation of  the propagator on the disc with mixed Neumann-Dirichlet
boundary conditions. 
Moreover, we compute the commutator of the coordinate fields 
confirming the well known results in the literature. 
In the final section we apply all our techniques to the explicit
construction of a ${\ZZ}_3\times \widehat{\ZZ}_3$ orientifold containing
D-branes with mixed boundary conditions.

\newsec{D-branes in asymmetric orbifolds}

In this section we investigate in which way open strings manage
to implement asymmetric symmetries. Naively, one might think
that asymmetric symmetries are an issue only
in the closed string sector, as open strings can be obtained by
projecting onto the left-right symmetric part of the space-time. 
However, historically just requiring  the asymmetric symmetry under T-duality
on a circle led  to the discovery
of D-branes. This T-duality acts on the space-time coordinates  as
\eqn\tdual{ (X_L,X_R)\to (-X_L,X_R) .} 
Thus, the open string sector deals with T-duality by giving rise
to a new kind of boundary condition leading in this case  to 
the well known Dirichlet boundary condition.  
Compactifying on a higher dimensional torus $T^d$, in general with 
non-zero $B$-fields, the T-duality
group gets  enlarged, so that one may ask what the image of Neumann boundary
conditions under these actions actually is. 
\pano
In the course of this paper we restrict ourselves to the two-dimensional
torus $T^2$ and direct products thereof. For concreteness consider
type IIB compactified on a $T^2$ with complex coordinate
$Z=X_1+iX_2$ allowing a discrete $\ZZ_N$ symmetry acting as
\eqn\symm{ \Theta: (Z_L,Z_R)\to \left(e^{i\theta} Z_L,
             e^{i\theta} Z_R\right) }
with $\theta=2\pi /N$.  
The essential observation is that performing a usual T-duality operation
in the $x_1$-direction 
\eqn\xdual{ T: (Z_L,Z_R)\to (-\o{Z}_L,Z_R)}
yields an asymmetric action on the T-dual torus $\hat{T}^2$
\eqn\xdualb{  \hat\Theta=T\Theta T^{-1}: (Z_L,Z_R)\to 
       \left(e^{-i\theta} Z_L,e^{i\theta} Z_R\right) .}
The aim of this paper is to investigate the properties of 
asymmetric orbifolds defined by actions like \xdualb.\pano
\vbox{
\diagram
 {\rm TYPE\,\, IIA} & \rTo^{R\leftrightarrow {\alpha^\prime / R}} 
                                           &   {\rm TYPE\,\, IIB} \\
 \dTo^{\hat\Theta}     &       &  \dTo_{\Theta}     \\
 {{\rm TYPE\,\, IIA} / \widehat{\ZZ}_N}& \rTo_{R\leftrightarrow 
 {\alpha^\prime /
R}} 
                                &   {{\rm TYPE\,\, IIB} / \ZZ_N}
\enddiagram 
\centerline{{\bf Diagram 1:} Duality relation}
}
\meno
\noindent
The strategy we will follow is  depicted in the commuting diagram 
1. In order to obtain the features of the asymmetric orbifold, 
concerning some questions it is appropriate to directly apply
the asymmetric rotation $\hat\Theta$. For other questions
it turns that it is better to first apply a T-duality and then
perform the symmetric rotation $\Theta$ and translate the result back
via a second T-duality. If not explictly present in the
equations, we have set $\alpha'=1$ both for the closed and the
open string.

\subsec{Definition of the T-dual torus}
\noindent
The first step is to define the T-dual torus $\hat{T}^2$ allowing indeed
an asymmetric action \xdualb. 
Let the torus $T^2$ be  defined by the following two
vectors
\eqn\vect{ e_1=R_1, \quad\quad e_2=R_2\, e^{i\alpha}, }
so that the complex and K\"ahler structures are given by
\eqn\kahl{\eqalign{  U&={e_2\over e_1}= {R_2\over R_1}\, e^{i\alpha}, \cr
                     T&=b+i R_1 R_2 \sin\alpha. \cr}}
The left and right moving zero-modes, i.e. Kaluza-Klein and winding
modes, can be written in the following form
\eqn\kkw{\eqalign{  p_L={1\over i\sqrt{U_2 T_2}} \left[ U\, m_1-m_2-\o{T}(n_1+
   U\, n_2) \right] ,\cr
       p_R={1\over i\sqrt{U_2 T_2}} \left[ U\, m_1-m_2- {T}(n_1+
   U\, n_2) \right]. \cr}} 
Applying T-duality in the $x_1$-direction exchanges the complex and 
the K\"ahler modulus yielding the torus $\hat{T}^2$ defined
by the vectors
\eqn\vectb{ \hat{e}_1={1\over R_1}, \quad\quad 
           \hat{e}_2={b\over R_1} + i R_2 \sin\alpha  }
and the two-form flux 
\eqn\twoform{ \hat{b}={R_2\over R_1} \cos\alpha.}
For the Kaluza-Klein and winding modes we get
\eqn\kkwb{\eqalign{  p_L&=-{1\over i\sqrt{\hat{U}_2 \hat{T}_2}} 
\left[ \hat{U}\, n_1+m_2-\hat{\o{T}}(m_1+
   \hat{U}\, n_2) \right], \cr
       p_R&=-{1\over i\sqrt{\hat{U}_2 \hat{T}_2}} \left[ \hat{U}\, n_1+m_2-
          {\hat{T}}(m_1+ \hat{U}\, n_2) \right], \cr}}
from which we deduce the relation of the Kaluza-Klein
 and winding quantum numbers
\eqn\rel{ \widehat{m}_1=-n_1,\ \ \widehat{m}_2=m_2,\ \ \widehat{n}_1=-m_1,\ \
             \widehat{n}_2=n_2.}
If the original lattice of  $T^2$ allows a crystallographic action of a
$\ZZ_N$ symmetry, then the T-dual Narain-lattice of
$\hat{T}^2$ does allow
a crystallographic action of the corresponding asymmetric $\widehat{\ZZ}_N$
symmetry. In view of the orientifold model studied in section 4, 
we present the $\ZZ_3$ case as an easy example. 

\subsec{The $\widehat{\ZZ}_3$ torus}
\noindent
One starts with the  $\ZZ_3$ lattice defined by the basis 
vectors
\eqn\vectzdrei{ e^{\bf A}_1=R, \quad\quad 
e^{\bf A}_2=R\left( {1\over 2}+i{\sqrt{3}\over 2}\right) }
and arbitrary $b$-field.
The complex and K\"ahler moduli are
\eqn\kahl{\eqalign{  U^{\bf A}&={1\over 2}+i{\sqrt{3}\over 2}, \cr
                     T^{\bf A}&=b+i R^2 {\sqrt{3}\over 2}. \cr}}
This lattice has the additional property that
it allows a crystallographic action of the reflection at the $x_2$-axis, 
${\cal R}$. This was important for the study of $\Omega{\cal R}$ orientifolds
in \rbgka. We call this lattice of type ${\bf A}$.
Recall from \rbgka, that under $\Omega {\cal R}$ all three $\ZZ_3$ fixed points
are left invariant. 
For zero $b$-field one obtains for instance for the T-dual ${\bf A}$ lattice
\eqn\vectzdreidual{ \hat{e}^{\bf A}_1={1\over R}, \quad\quad 
       \hat{e}^{\bf A}_2=iR{\sqrt{3}\over 2}}
and $\hat{b}^{\bf A}={1/2}$.
That this rectangular lattice 
features an asymmetric $\widehat{\ZZ}_3$ symmetry and that all three 
``fixed points'' of the $\widehat{\ZZ}_3$ are left invariant under $\Omega$
is not obvious at all. This  shows already how
T-duality can give rise to fairly non-trivial results.
\pano
As we have already shown in \rbgka\ there exists a second $\ZZ_3$ lattice, 
called type ${\bf B}$, 
allowing a crystallographic action of the reflection ${\cal R}$.
The basis vectors are given by
\eqn\vectzdreib{ e^{\bf B}_1=R, \quad\quad 
            e^{\bf B}_2={R\over 2}+i{R\over 2\sqrt{3}} }
with arbitrary $b$-field leading to the  complex and K\"ahler moduli
\eqn\kahlb{\eqalign{  U^{\bf B}&={1\over 2}+i{1\over 2\sqrt{3}},  \cr
                     T^{\bf B}&=b+i {R^2 \over 2 \sqrt{3} }. \cr}}
For the ${\bf B}$ lattice only one  $\ZZ_3$ fixed point is invariant
under $\Omega {\cal R}$, the remaining two are interchanged. 
For $b=0$ the T-dual lattice is defined by 
\eqn\vectzdreidual{ \hat{e}^{\bf B}_1={1\over R}, \quad\quad 
       \hat{e}^{\bf B}_2=i{R\over 2\sqrt{3}}}
with  $\hat{b}^{\bf B}={1/2}$.
It is a non-trivial  consequence of T-duality that only one of the three 
$\widehat{\ZZ}_3$ ``fixed points'' is left invariant under
$\Omega$.
\pano
If one requires the lattices to allow simultaneously  a symmetric $\ZZ_3$
and an asymmetric
$\widehat{\ZZ}_3$ action one is stuck at the self-dual point $U=T$ yielding
$R=1$ and $b=1/2$. Note, that this is precisely  the root lattice of the 
$SU(3)$
Lie algebra. Since now we are equipped with lattices indeed allowing
a crystallographic action of asymmetric $\widehat{\ZZ}_N$ operations,
we can move forward to discuss their D-brane contents.

\subsec{Asymmetric rotations of D-branes}
\noindent
In order to divide a string theory by some discrete group we first have
to make sure that the theory is indeed invariant. For the open 
string sector this means that the D-branes also have to be arranged in such a
way that they reflect the discrete symmetry.
Thus, for instance we would like to know what the image of a D0-brane
under an asymmetric
rotation is. In the compact case we can ask this question for
the discrete $\widehat{\ZZ}_N$ rotations defined in the last subsection,
but we can also pose it quite generally in the non-compact case
using a continuous  asymmetric rotation
\bigno
\eqn\rot{ \left(\matrix{ X'_{1,L} \cr X'_{2,L} \cr} \right)=\left(
           \matrix{ \cos\theta & \sin\theta \cr 
                    -\sin\theta & \cos\theta \cr}\right) 
             \left(\matrix{ X_{1,L} \cr X_{2,L} \cr} \right), \quad\quad
           \left(\matrix{ X'_{1,R} \cr X'_{2,R} \cr} \right)=\left(
           \matrix{ \cos\theta & -\sin\theta \cr 
                    \sin\theta & \cos\theta \cr}\right) 
             \left(\matrix{ X_{1,R} \cr X_{2,R} \cr} \right) .} 
As outlined already in the beginning of section 2 (see diagram 1), 
instead of acting with the asymmetric rotation on the Dirichlet boundary 
conditions of the D0-brane, it is equivalent to go to the T-dual picture,
apply first a symmetric rotation on the branes and then perform a
T-duality transformation in the $x_1$-direction.             
In the T-dual picture the D0-brane becomes a D1-brane filling only the 
$x_1$-direction. 
Thus, the open strings are of Neumann type in the $x_1$-direction and
of Dirichlet type in the $x_2$-direction.
The asymmetric rotation becomes a symmetric rotation, which simply
rotates the D1-brane by an angle $\theta$ in the $x_1$-$x_2$ plane.   
Thus, after the rotation the D1 boundary conditions in these two directions
read 
\smno
\eqn\bounda{\eqalign{  \partial_\sigma X_1 + \tan\theta\, 
        \partial_\sigma X_2 &=0, \cr
             \partial_\tau X_2 - \tan\theta\, \partial_\tau X_1 &=0. \cr}}
If we are on the torus $T^2$ there is a distinction between values
of $\theta$, for which the rotated D1-brane intersects a lattice point, and
values of $\theta$, for which the D1-brane  densely covers the entire
$T^2$. In the first case, one still obtains  
quantized Kaluza-Klein and winding modes
as computed in \raas. 
\pano
If the D1-brane runs $n$-times around the $e_1$ circle
and $m$ times around the $e_2$ circle until it intersects a lattice point, 
the relation
\eqn\inte{  \cot\theta=\cot\alpha+{n\over m U_2}}
holds. As an example we show in figure 1 a rotated D1-brane with $n=2$ and
$m=1$.

\fig{}{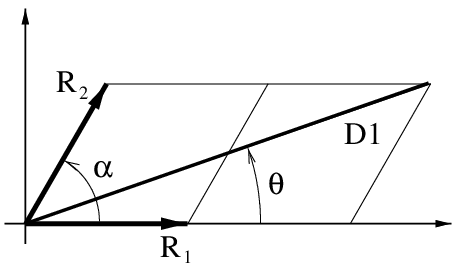}{8truecm}
\noindent
In the following we will mostly consider D-branes of the
first kind, which we will call rational D-branes.  
Finally, T-duality in the $x_1$-direction has the effect of exchanging 
$\partial_\sigma X_1\leftrightarrow  - \partial_\tau X_1$, leading to the
boundary conditions \rjabbari
\eqn\boundb{\eqalign{  \partial_\sigma X_1 + \cot\theta\, 
              \partial_\tau X_2 &=0 ,\cr
             \partial_\sigma X_2 - \cot\theta\, \partial_\tau X_1 &=0 .\cr}}
As emphasized already, 
one could also perform the asymmetric rotation directly on the Dirichlet
boundary conditions for the D0-brane and derive the same result. Thus, 
we conclude that an asymmetric rotation turns a D0-brane into 
a D2-brane with mixed boundary conditions. The last  statement is the main
result of this paper. As has been discussed intensively in the 
last year, mixed boundary conditions arise from open strings
travelling in a background with non-trivial two-form flux, $B$,  or
non-trivial gauge flux, $F$,
\eqn\boundc{\eqalign{  \partial_\sigma X_1 + (B+F)\, \partial_\tau X_2 &=0, \cr
                  \partial_\sigma X_2 - (B+F)\, \partial_\tau X_1 &=0 .\cr}}
Thus, we can generally identify 
\eqn\ident{  \cot\theta={\cal F}={B+F},}
which in the rational case becomes (note that $\cot\theta$ is not necessarily
rational)
\eqn\identb{  \cot\theta=\cot\alpha + {n\over m U_2} ={B+F} .}
Since the $B$ field is related to the shape of the torus $T^2$ and
the $F$ field to the D-branes, from  \identb\ we extract the 
following identifications 
\eqn\schnee{ B=\cot\alpha, \quad\quad F={n\over m U_2}.}
In section 3  we will further elaborate the relation between
asymmetric rotations and D-branes with mixed boundary conditions and
will present an alternative derivation of some of the noncommutativity 
properties known  for such boundary conditions.
In the remainder of this section we will focus our attention on  the
zero mode spectrum for open strings stretched between D-branes with mixed 
boundary conditions.
In particular, we will demonstrate  that in the compact case   
open strings stretched between identical rational D-branes  do  
have a non-trivial zero mode spectrum.
This is in sharp contrast to some statements in the literature \rkaku\
saying that Neumann boundary conditions allow Kaluza-Klein 
 momentum, Dirichlet boundary
conditions allow non-trivial winding but general mixed D-branes do have
neither of them.  

\subsec{Kaluza-Klein and winding modes}
\noindent
Since we can not easily visualize a D-brane with mixed boundary conditions,
we first determine the zero-mode spectrum in the closed string
tree channel and then 
transform the result into the open string loop channel.
Thus, we are looking for boundary states (see also \rbrunner)
in the closed string theory
satisfying the following boundary state conditions
\eqn\boundd{\eqalign{ 
 \left[\partial_\tau X_{1,cl} + \cot\theta\, \partial_\sigma
   X_{2,cl}\right]|B\rangle 
    &=0 ,\cr
  \left[\partial_\tau X_{2,cl} - \cot\theta\, \partial_\sigma
    X_{1,cl}\right]|B\rangle 
      &=0 .\cr}}
Rewriting \boundd\ in terms of the complex coordinate  
the boundary condition reads
\eqn\bounde{ \left[\partial_\tau Z_{cl} -i \cot\theta\, \partial_\sigma
   Z_{cl}\right]|B\rangle =0. }
Using the mode expansion
\eqn\mode{\eqalign{ Z_{cl}&={z_0\over 2} + {1\over 2}(p_L+p_R)\tau + 
{1\over 2} (p_L-p_R)\sigma + {i\over \sqrt{2}} 
      \sum_{n\ne 0} \left( {\alpha_n\over n} e^{-in (\tau+\sigma)} +
              {\widetilde\alpha_n\over n} e^{-in (\tau-\sigma)} \right), \cr
   \o{Z}_{cl}&={\o{z}_0\over 2} + {1\over 2}(\o{p}_L+\o{p}_R)\tau + 
{1\over 2} (\o{p}_L-\o{p}_R)\sigma -
 {i\over \sqrt{2}}\sum_{n\ne 0} \left( {\o{\alpha}_n\over n} 
     e^{-in (\tau+\sigma)} +
      {\o{\widetilde\alpha}_n\over n} e^{-in (\tau-\sigma)} \right) \cr}}
one obtains 
\eqn\modeb{\eqalign{\left[ (p_L+p_R)-i\cot\theta\,(p_L-p_R)\right]
       |B\rangle&=0 ,\cr
      \left[ \alpha_n+e^{2i\theta} \widetilde{\alpha}_{-n} \right]
      |B\rangle&=0\cr}}
with similar conditions for the fermionic modes. 
Inserting \kkwb\ and \rel\ into the first equation of \modeb\ one can solve
for the Kaluza-Klein and winding modes
\eqn\solve{ \widehat{m}_1=-{n\over m} \widehat{n}_2,\quad\quad 
            \widehat{m}_2={n\over m} \widehat{n}_1}
giving rise to the following zero-mode spectrum
\eqn\spec{  M^2_{cl}={ |r+s\,\hat{U}|^2\over \hat{U}_2}\, 
                     { |n+m\, \hat{T}|^2\over \hat{T}_2}}
with $r,s\in\ZZ$.
We observe that this agrees with the spectrum derived in \raas\ by 
employing T-duality. 
Note, that the formula \spec\ is explicitly $SL(2,\ZZ)\times SL(2,\ZZ)$ 
invariant. Thus, the bosonic part of a boundary state satisfying 
\modeb\ is given by
\eqn\boundsol{ |B\rangle_{(n,m)} =\sum_{r,s\in\ZZ} {\rm exp}
  \left(\sum_{n\in\ZZ}
  {1\over n} e^{2 i\theta} \alpha_{-n} {\widetilde{\alpha}}_{-n} \right) 
     |r,s\rangle_{(n,m)}. }
Using this boundary state we compute the tree channel annulus partition 
function. 
Transforming  the result via a modular transformation into the loop channel,
we can extract the zero-mode contribution and conclude that open
strings stretching between identical rational  D-branes 
carry non-vanishing zero modes giving rise to masses
\eqn\specb{  M^2_{o}={ |r+s\,\hat{U}|^2\over \hat{U}_2}\, 
                     {  \hat{T}_2 \over |n+m\,\hat{T}|^2} .}
It would be interesting to derive this quantization condition directly
in the open string sector. Again making use of T-duality in the
$x_1$-direction allows use to extract also the individual Kaluza-Klein and 
winding contributions
\eqn\indiv{\eqalign{ \hat{P}&={1\over 2}\left(\hat{p}_L+\hat{p}_R\right)
     =\sqrt{\hat{T}_2\over \hat{U}_2} 
  {m\hat{T}_2\over |n+m\,\hat{T}|^2}\left(s\,\hat{U}_2+i(r+s\,\hat{U}_1 )
\right),
  \cr
   \hat{L}&={1\over 2}\left( \hat{p}_L-\hat{p}_R\right)=-\sqrt{\hat{T}_2\over
      \hat{U}_2} 
  {n+m\hat{T}_1\over |n+m\,\hat{T}|^2}
  \left((r+s\,\hat{U}_1 )-is\,\hat{U}_2\right). \cr}}
Note, that the zero modes indeed satisfy the boundary condition \boundb.
Summarizing, we now have the means to compute annulus amplitudes
for open strings stretched between different kinds of D-branes with
rational mixed boundary conditions. 
As an example, we discuss the $\widehat{\ZZ}_3$ case
in some more detail.

\subsec{D-branes in the asymmetric $\widehat{\ZZ}_3$ orbifold}
\noindent
Consider the $\widehat{\ZZ}_3$ lattice of type ${\bf A}$ and start with
a D$_1$-brane with pure Dirichlet boundary conditions ($\theta=0$)
\eqn\branea{  \eqalign{  \partial_\tau X_1 &=0 ,\cr
                          \partial_\tau X_2 &=0 .\cr}}
Successively applying the asymmetric $\widehat{\ZZ}_3$ this D-brane is 
mapped to 
a mixed D$_2$-brane with boundary conditions ($\theta=2\pi/3$)
\eqn\braneb{\eqalign{  \partial_\sigma X_1 - 
                 {1\over \sqrt{3}}\partial_\tau X_2 &=0, \cr
                      \partial_\sigma X_2 + 
                    {1\over \sqrt{3}} \partial_\tau X_1 &=0 \cr}}
and a mixed D$_3$-brane with boundary conditions ($\theta=-2\pi/3$)
\eqn\braneb{\eqalign{  \partial_\sigma X_1 + {1\over \sqrt{3}}
             \partial_\tau X_2 &=0, \cr
                      \partial_\sigma X_2 - 
            {1\over \sqrt{3}} \partial_\tau X_1 &=0 .\cr}}
In the orbifold theory these three kinds of D-branes are identified. 
This reflects that their background fields are being identified according to 
\eqn\ident{ {\cal F} \equiv {\cal F} + {1 \over \sqrt{3}} }
or equivalently 
\eqn\identb{ \theta \equiv \theta + {2\pi \over 3}.} 
The two coordinates $X_1$ and $X_2$ yield the following 
contribution to the annulus partition function for open strings 
stretched between identical D-branes
\eqn\anna{ A^{\alpha\beta}_{ii}= {\th{\alpha}{\beta}\over \eta^3}
     \left(\sum_{r\in\ZZ} e^{-2\pi t{r^2\over R^2}}\right) 
     \left(\sum_{s\in\ZZ} e^{-2\pi t{3s^2\over 4R^2}}\right) }
independent of  $i\in\{1,2,3\}$. 
Open strings stretched between different kinds of
D-branes give rise to shifted moding  and yield the partition
function
\eqn\annb{ A_{i,i+1}=n_{i,i+1} {\th{{1\over 3}+\alpha}{\beta}
        \over \th{{1\over 3}+\alpha}{{1\over 2}} } }
which looks like a twisted open string sector.
As we know from \refs{\rbgka\rba-\rbgkb} here we have to take into
account extra multiplicities, $n_{i,i+1}$, which have a natural
geometric interpretation as  multiple intersection
points of D-branes at angles  in the T-dual picture.
By this reasoning we find that
for the {\bf A} type lattice the extra factor is one. 
However, for the three D-branes generated by $\widehat{\ZZ}_3$ 
when one starts with a D-brane with pure Neumann boundary conditions,
$\theta\in\{\pi/2,\pi/6,-\pi/6\}$, 
T-duality tells us that there must appear an extra factor of three
in front of the corresponding annulus amplitude 
\annb. In the orientifold construction presented in section 4
these multiplicities are important to give consistent models. 
    
\newsec{Asymmetric rotations and noncommutative geometry}

In section 2 we have pointed out that on $T^2$ or $\IR^2$ D-branes with 
mixed boundary
conditions can be generated by simply applying an asymmetric
rotation to an ordinary D-brane with pure Neumann or Dirichlet
boundary conditions. Thus, it should be possible to rederive earlier
results for the two-point function on the disc
\eqn\twopoint{  \langle X_i(z)\, X_j(z') \rangle, }
for the operator product expansion (OPE) between   vertex operators
on the boundary
\eqn\vertb{ e^{i p X}(\tau)\,   e^{i q X}(\tau') }
or for the commutator of the coordinate fields 
\eqn\comm{  [ X_i(\tau,\sigma), X_j(\tau,\sigma') ] }
by applying an asymmetric rotation on the corresponding quantities
for open strings ending on D0-branes in flat space-time.

\subsec{Two-point function on the disc}
\pano
The two-point function on the disc for both $X_1$ and $X_2$ of Dirichlet type
reads
\eqn\twopointb{\eqalign{  \langle X_i(z)\, X_j(z') \rangle&=
                 -\alpha'\delta_{ij}\left( \ln |z-z'| -\ln |z-\o{z}'| \right)
                 \cr
                &=-\alpha'\delta_{ij}{1\over 2}
                \left( \ln (z-z') + \ln (\o{z}-\o{z}')
             -\ln (z-\o{z}') - \ln (\o{z}-{z}')\right) \cr }}
from which, formally using 
\eqn\move{ X_i(z)=X_{i,L}(z) + X_{i,R}(\o{z}), }
we can directly read off the individual contributions from
the left- and right-movers.
Performing the asymmetric rotation
\eqn\asymrota{ X_L\to A\, X_L, \quad\quad X_R\to A^T X_R ,}
where $A$ denotes an element of $SO(2)$, leads to the 
following expression
for the propagator in the rotated coordinates
\eqn\twopointb{\eqalign{  \langle X_i(z)\, X_j(z') \rangle=&
                 -\alpha'\delta_{ij} \ln |z-z'| - \alpha'\delta_{ij}
               \left( \sin^2\theta-\cos^2\theta\right) \ln |z-\o{z}'| -\cr
               &\alpha' \epsilon_{ij} \sin\theta\, \cos\theta\,
               \ln\left({z-\o{z}'\over \o{z}-z' } \right) .}}
This expression agrees precisely  with the propagator derived in \rnappi\
with the identification
\eqn\nappi{   {\cal F}=\left(\matrix{ 0 & \cot\theta \cr
                               -\cot\theta & 0 \cr} \right) .}
Thus, by applying an asymmetric rotation we have found an elegant and short
way of deriving this  propagator without explicit reference to the
boundary conditions or the background fields. Moreover, since the
commutative D0-brane is related in this
smooth way to a noncommutative  D2-brane, it is suggesting that
also both effective theories arising on such branes are related
by some smooth 
transformation. Such an  explicit map between the commuting and
the noncommuting effective gauge theories  has been determined in \rseiwit.

\subsec{The OPE of vertex operators}
\pano
In this subsection we apply an asymmetric rotation also
to the operator product expansion of tachyon vertex operators 
${\cal O}(z)= e^{ipX}(z)$ on the 
boundary. Of course this OPE is a direct consequence of the
correlator \twopointb\ restricted to the boundary, but nevertheless
we would like to see whether we can generate  the noncommutative
$\ast$-product directly
via an asymmetric rotation. 
Taking care of the left- and right-moving contributions in the
OPE between vertex operators living on a pure Dirichlet boundary
we can write for $|z|>|z'|$
\eqn\verta{ e^{i p X}(z)\,   e^{i q X}(z') =
        {(z-z')^{{\alpha'\over 2} p_L q_L }  \,
        (\o{z}-\o{z}')^{{\alpha'\over 2} p_R q_R }
       \over (z-\o{z}')^{{\alpha'\over 2} p_L q_R }\,
             (\o{z}-{z}')^{{\alpha'\over 2} p_R q_L }}\
             e^{i (p+q) X}(z')+\ldots .}
Now we apply an asymmetric rotation \asymrota\ together with 
\eqn\asymrot{\eqalign{ &p_L\to A\, p_L, \quad\quad p_R\to A^T p_R, \cr
                       &q_L\to A\, q_L, \quad\quad q_R\to A^T q_R, \cr}}
and,  after all,  identifying
$p_L=p_R$, $q_L=q_R$ we obtain
\eqn\vertba{  e^{i p X}(z)\,   e^{i q X}(z') =
           {\left[(z-z')  
        (\o{z}-\o{z}')\right]^{{\alpha'\over 2} p q } \over
        \left[ (z-\o{z}')
             (\o{z}-{z}')\right]^{{\alpha'\over 2} \cos(2\theta)\, p q }}
         \left( {z-\o{z}'\over \o{z}-z'} \right)^{-{\alpha'\over 2}
        \epsilon_{ij} p_i q_j \sin(2\theta) } \
             e^{i (p+q) X}(z')+\ldots }
Restricting \vertba\ to the boundary and choosing the same branch cut as 
in \rseiwit\ we finally arrive at 
\eqn\vertb{  e^{i p X}(\tau)\,   e^{i q X}(\tau') =
           (\tau-\tau')^{\alpha' p q (1+\sin^2\theta-\cos^2\theta)}\,
            {\rm exp}\left(-i \pi \alpha' \sin\theta \cos\theta  
           \epsilon_{ij} p_i q_j\right) \ 
             e^{i (p+q) X}(\tau')+\ldots .}
This is precisely the OPE derived in 
\refs{\rschomi,\rseiwit}. It shows that it is indeed possible to derive the 
$\ast$-product $e^{ipX}(\tau)e^{iqX}(\tau')\sim e^{ipX}\ast e^{iqX}(\tau')$
directly via an asymmetric rotation, 
where the noncommutative algebra ${\cal A}$ of functions $f$ and $g$
is defined as
\eqn\starprod{
f\ast g=fg-i\pi \alpha' \sin\theta\, \cos\theta\, \epsilon_{ij}\,
\partial_if\partial _jg+ \dots.}

\subsec{The commutator of the coordinates}
\pano
While the two-point function derived above already implies that the commutator of 
the coordinate fields is non-vanishing, i.e. the geometry on the D-brane 
non-commutative, we would like to rederive this result directly via 
studying D-branes with mixed boundary conditions, as well. 
This is done by the quantization of the bosonic coordinate fields of 
the open string. We start with the T-dual situation
with two D-branes intersecting at an arbitrary angle $\theta_2-\theta_1$ 
(see figure 2).

\fig{D-Branes at angles}{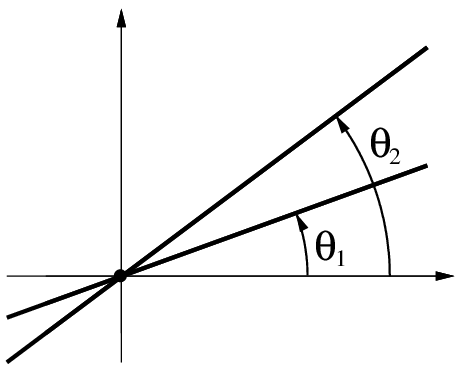}{5truecm}
\noindent
The open string boundary condition at $\sigma=0$ are 
\eqn\sigmanull{\eqalign{  \partial_\sigma X_1 + \tan\theta_1\, 
                   \partial_\sigma X_2 &=0, \cr
                      \partial_\tau X_2 - \tan\theta_1\, \partial_\tau X_1
                      &=0, \cr}}
and at $\sigma=\pi$  we require
\eqn\sigmanullb{\eqalign{  \partial_\sigma X_1 + \tan\theta_2\, 
                   \partial_\sigma X_2, &=0 ,\cr
                      \partial_\tau X_2 - \tan\theta_2\, \partial_\tau X_1
                      &=0. \cr}}
The mode expansion satisfying these two boundary conditions looks like
\eqn\modeg{\eqalign{ X_1=x_1 +&
    i \sqrt{\alpha'} \sum_{n\in\ZZ}  {\alpha_{n+\nu}\over n+\nu} 
           e^{-i(n+\nu) \tau} \cos[(n+\nu)\sigma +\theta_1] + \cr
    &i \sqrt{\alpha'} \sum_{m\in\ZZ}  {\alpha_{m-\nu}\over m-\nu} 
           e^{-i(m-\nu) \tau} \cos[(m-\nu)\sigma -\theta_1], \cr
    X_2=x_2 +&
    i \sqrt{\alpha'} \sum_{n\in\ZZ}  {\alpha_{n+\nu}\over n+\nu} 
           e^{-i(n+\nu) \tau} \sin[(n+\nu)\sigma +\theta_1] -\cr
    &i \sqrt{\alpha'} \sum_{m\in\ZZ}  {\alpha_{m-\nu}\over m-\nu} 
           e^{-i(m-\nu) \tau} \sin[(m-\nu)\sigma -\theta_1], \cr }}
with $\nu=(\theta_2-\theta_1)/\pi$. Using the usual commutation relation 
\eqn\osci{  [\alpha_{n+\nu}, \alpha_{m-\nu}]=(n+\nu)\, \delta_{m+n,0} }
and the vanishing of the commutator of the center of mass coordinates $x_1$ 
and $x_2$ one can easily show that for D-branes at angles the 
general equal time commutator vanishes
\eqn\commc{  [ X_i(\tau,\sigma), X_j(\tau,\sigma') ]=0 .}
Therefore, the geometry of D-branes at angles, but without background 
gauge fields, is always commutative.  

Performing  T-duality in the $x_1$ direction one gets the two 
mixed boundary conditions for the open strings
\eqn\sigmanullc{\eqalign{  \partial_\sigma X_1 + \cot\theta_1\, 
                   \partial_\tau X_2 &=0 ,\cr
                      \partial_\sigma X_2 - \cot\theta_1\, \partial_\tau X_1
                      &=0 \cr}}
at $\sigma=0$ and 
\eqn\sigmanulld{\eqalign{  \partial_\sigma X_1 + \cot\theta_2\, 
                   \partial_\tau X_2 &=0, \cr
                      \partial_\sigma X_2 - \cot\theta_2\, \partial_\tau X_1
                      &=0 \cr}}
at $\sigma=\pi$.
The mode expansion satisfying these boundary conditions is
\eqn\modeh{\eqalign{ X_1=x_1 -&
    \sqrt{\alpha'} \sum_{n\in\ZZ}  {\alpha_{n+\nu}\over n+\nu} 
           e^{-i(n+\nu) \tau} \sin[(n+\nu)\sigma +\theta_1] - \cr
    &\sqrt{\alpha'} \sum_{m\in\ZZ}  {\alpha_{m-\nu}\over m-\nu} 
           e^{-i(m-\nu) \tau} \sin[(m-\nu)\sigma -\theta_1], \cr
    X_2=x_2 +&
    i \sqrt{\alpha'} \sum_{n\in\ZZ}  {\alpha_{n+\nu}\over n+\nu} 
           e^{-i(n+\nu) \tau} \sin[(n+\nu)\sigma +\theta_1] -\cr
    &i \sqrt{\alpha'} \sum_{m\in\ZZ}  {\alpha_{m-\nu}\over m-\nu} 
           e^{-i(m-\nu) \tau} \sin[(m-\nu)\sigma -\theta_1], \cr }}
which one can also derive performing an asymmetric rotation
on the mode expansion for pure Dirichlet type branes. 
Now one can compute the commutator
\eqn\commd{  [ X_1(\tau,\sigma), X_2(\tau,\sigma') ]=[x_1,x_2]+
                   i\sum_{n\in\ZZ} {2\alpha' \over n+\nu} \sin[(n+\nu)\sigma
                   +\theta_1] \sin[(n+\nu)\sigma' +\theta_1] .}
For $\sigma$ and $\sigma'$ not both equal to zero or  $\pi$ the second term
in \commd\ is constant, so that the whole expression can be set to zero
by choosing 
\eqn\choose{  [x_1,x_2]={ 2 \pi i \alpha' \over {\cal F}_2-{\cal F}_1} .}
Note, that this value  also can be obtained  directly
from the canonical quantization \rnappi.
However, as has been shown in \refs{\rchen,\rchu}, for $\sigma=\sigma'=0$
the evaluation of the sum in \commd\ yields
\eqn\comme{  [ X_1(\tau,0), X_2(\tau,0) ]=-{2\pi i \alpha' {\cal F}_1\over 
      1+{\cal F}_1^2}}
and for $\sigma=\sigma'=\pi$ one analogously obtains
\eqn\commf{  [ X_1(\tau,\pi), X_2(\tau,\pi) ]={2 \pi i\alpha' {\cal F}_2\over 
1+{\cal F}_2^2}.}
Thus, the  coordinates only noncommute at the boundary of the word-sheet, 
where the commutator can be expressed entirely in terms of the gauge field 
on the local D-brane. 
If both ends of the open string end on the same D-brane with 
$\theta_1=\theta_2$, both terms in \commd\ become singular, but the
sum of them give rise to the same expression \comme\ and \commf\
for the commutator at the ends of the open string. Thus, only for
$\theta_1=\theta_2\in\{0,\pi/2\}$ the $X_1$ and $X_2$ coordinates commute on the
D-brane, in all other cases the end points see a noncommuting
space-time. 
Moreover, in the compact case for rational D-branes
the noncommutative theory on the D-branes
is mapped via T-duality to a commutative theory on D-branes at angles. 
This is only a special example
of the more general rule pointed out in \rseiwit\ 
that for rational points the noncommutative torus is T-dual to a commutative 
one.

At the end of this section let us briefly comment on the algebraic  
structure of the noncommutative torus we have obtained by the asymmetric
rotation on the D-branes. As shown in the previous section, the tachyon
vertex operator ${\cal O}=e^{ipX}(\tau)$ leads to a noncommutative
algebra ${\cal A}$, defined in eq.\starprod.
As explained in \rseiwit, the algebra ${\cal A}$ of tachyon vertex operators
can be taken at either end of the open string.
Therefore the open string states form a bimodule ${\cal A}\times {\cal A}'$,
where ${\cal A}$ is acting on the boundary $\sigma=0$ and
${\cal A}'$ on the boundary $\sigma=\pi$ of the open string.
Specifically, for an open string whose first boundary $\sigma =0$ is 
related to
a D-brane with parameter $\theta_1$ and whose second boundary $\sigma=\pi$
is attached to a D-brane with parameter $\theta_2$, the algebra ${\cal A}$
of functions on the noncommutative torus
is generated
by 
\eqn\ua{\eqalign{
&U_1=\exp(iy_1-{2\pi^2 \alpha' 
{\cal F}_1\over 1+{\cal F}_1^2}(\partial /\partial y_2)),
\cr 
&U_2=\exp(iy_2+{2\pi^2 \alpha'
{\cal F}_1\over 1+{\cal F}_1^2}(\partial /\partial y_1)),
\cr }}
which obey 
\eqn\uu{U_1U_2=\exp(-2\pi i{2\pi\alpha'{\cal F}_1\over 1+{\cal F}_1^2})U_2U_1.}
On the other hand, the algebra 
 ${\cal A}'$
is generated
by 
\eqn\utilde{\eqalign{
&\tilde U_1
=\exp(iy_1+{2\pi^2\alpha'{\cal F}_2\over 1+{\cal F}_2^2}
(\partial /\partial y_2)),\cr
&\tilde U_2
=\exp(iy_2-{2\pi^2\alpha'{\cal F}_2\over 1+{\cal F}_2^2}
(\partial /\partial y_1)), \cr}}
obeying 
\eqn\uutilde{\tilde U_1\tilde U_2
=\exp(2\pi i{2\pi\alpha'{\cal F}_2\over 1+{\cal F}_2^2})\tilde U_2\tilde U_1.}

\newsec{Asymmetric orientifolds}

Another motivation for studying such asymmetric orbifolds arises in 
the construction of type I vacua. In \refs{\rbgka\rba-\rbgkb}
we have 
considered so-called supersymmetric 
orientifolds with D-branes at angles in six and four space-time dimensions 
which in the six-dimensional case were defined as
\eqn\oriend{ {{\rm Type\, IIB}\ {\rm on}\ T^4\over 
       \{\Omega{\cal R}, \Theta \} }}
with ${\cal R}:z_i\to -\o{z}_i$, the $z_i$ being the complex coordinates of the $T^4$. 
Upon T-dualities in the directions of their real parts 
one obtains an ordinary orientifold where, however,
the space-time symmetry becomes asymmetric
\eqn\orienc{ {{\rm Type\, IIB}\ {\rm on}\ \hat{T}^4\over 
   \{\Omega, \hat\Theta \} }. }
In the entire derivation in section 2 we 
have identified the two constructions explicitly via T-duality, relating 
branes with background fields to branes at angles. 
While in the $\Omega {\cal R}$ orientifolds $\Theta$ identfied branes at different 
locations on the tori, $\hat{\Theta}$ now maps branes with different values 
of their background gauge flux upon each other. As the background fields determine 
the parameter which rules the noncommutative geometry, branes with different 
geometries are identified according to \ident. In this manner 
asymmetric orbifolds and orientifolds provide extremely exotic string backgrounds. 
In the particular example of the $\widehat{\ZZ}_3$ orientifolds, as obtained from the 
$\ZZ_3$ $\Omega {\cal R}$ orientifold in \rbgka\ via T-duality, all background fields 
are equivalent to a vanishing field, all geometries equivalent to a commutative one. 
But in any of the models, where the orbifold group contains an element of order 2, 
i.e. with even $N$,  
the background fields can only be ``gauged away'' on one half of the D-branes, 
the other half stays noncommutative.   

From the above mentioned identification it is now clear that the 
${\cal N}=(0,1)$ supersymmetric asymmetric $\widehat{\ZZ}_N$ orientifolds
\orienc\ have the same one loop partition functions as the corresponding 
symmetric  $\ZZ_N$ orientifolds \oriend.
The only difference is that instead of D7-branes at angles, we introduce
D9-branes with appropriate background fields. 
Thus, a whole class of asymmetric orientifolds has already been
studied in the T-dual picture involving D-branes at angles. 
One could repeat the whole computation for
the asymmetric orientifolds \orienc, getting of course identical results. 
Note, the model \orienc\ is really a type I vacuum, as $\Omega$
itself is gauged. Thus, in principle there exist the possibility
that heterotic dual models exist. Of course, in six dimensions most models
have more than one tensormultiplet so that no perturbative heterotic
dual model can exist. It would be interesting to look for 
heterotic duals for the four dimensional models discussed in \rbgkb.
\pano
In the following we will construct the even more general 
six-dimensional $\ZZ_3\times
\widehat{\ZZ}_3$ orientifold
\eqn\orienc{ {{\rm Type\, IIB}\ {\rm on}\ \hat{T}^4\over 
            \{\Omega, \Theta,\hat\Theta \}} }
which is T-dual to 
\eqn\oriend{ {{\rm Type\, IIB}\ {\rm on}\ T^4\over \{\Omega{\cal R},\hat\Theta,
\Theta \} },}
where in fact, as shown in section 2.2, the two tori are identical
$T^4=\hat{T}^4=SU(3)^2$. The freedom to choose their complex structures 
gives rise 
to a variety of three distinct models, which are denoted by ${\bf AA,AB,BB}$ as in 
\rbgka. 
Note, that the same orbifold group is generated
by a pure left-moving $\ZZ_{3L}$, $\Theta_L=\hat\Theta\Theta^{-1}$, and
a  pure right-moving $\ZZ_{3R}$, $\Theta_R=\hat\Theta\Theta$.
As was also shown in \rbimopra\ this model actually
has ${\cal N}=(1,1)$ supersymmetry, but one can get 
${\cal N}=(0,1)$ supersymmetry by turning on non-trivial discrete
torsion. 

\subsec{Tadpole cancellation}
\noindent
The computation of the various one-loop amplitudes is straightforward.
For the loop channel Klein bottle amplitude  we obtain
\eqn\klein{\eqalign{ K^{(ab)}={8c\over 12} \int_0^\infty {dt\over t^4} 
      \biggl[ &\rho_{00}\, \Lambda^a \Lambda^b + \rho_{01} + \rho_{02} +\cr
            & n^{(ab)}_{\hat\Theta,\Omega}\, \rho_{10}+
              n^{(ab)}_{\hat\Theta,\Omega\Theta^2}\,\epsilon \rho_{11}+
             n^{(ab)}_{\hat\Theta,\Omega\Theta}\,\o\epsilon \rho_{12}+\cr
            & n^{(ab)}_{\hat\Theta^2,\Omega}\, \rho_{20}+
              n^{(ab)}_{\hat\Theta^2,\Omega\Theta^2}\,\o\epsilon \rho_{21}+
          n^{(ab)}_{\hat\Theta^2,\Omega\Theta}\,\epsilon \rho_{22} \biggr],}}
where $c\equiv {\rm V}_6/\left( 8\pi^2 \alpha^\prime \right)^3$ and $\epsilon$ is 
a phase factor defining the discrete torsion. Further we 
have introduced a similar  notation as in \rbimopra\
\eqn\rhos{\eqalign{ \rho_{00}&=\sum_{\alpha,\beta=0,{1\over 2}}
           (-1)^{2\alpha+2\beta+4\alpha\beta}{\th{\alpha}{\beta}^4\over 
                \eta^{12}}, \cr 
            \rho_{0h}&=\sum_{\alpha,\beta=0,{1\over 2}}
           (-1)^{2\alpha+2\beta+4\alpha\beta}{\th{\alpha}{\beta}^2\over 
                \eta^6} \prod_{i=1}^2 2\sin(\pi h_i)
                {\th{\alpha}{\beta+h_i}\over 
                \th{{1\over 2}}{{1\over 2}+h_i}},\quad h\ne 0, \cr
            \rho_{gh}&=\sum_{\alpha,\beta=0,{1\over 2}}
           (-1)^{2\alpha+2\beta+4\alpha\beta}{\th{\alpha}{\beta}^2\over 
                \eta^6} \prod_{i=1}^2 
                {\th{\alpha+g_i}{\beta+h_i}\over 
                \th{{1\over 2}+g_i}{{1\over 2}+h_i}},\quad g,h\ne 0 \cr}}
with $g,h\in\{(1/3,-1/3),(2/3,-2/3)\}$ for which we use the shorter
notation $g,h\in\{1,2\}$.
The index $(ab)$ denotes the three possible choices of lattices,
${\bf AA}$, ${\bf AB}$ and ${\bf BB}$, and $\Lambda^a$ are the zero mode
contributions \specb\ to the partition function
\eqn\zeros{\eqalign{ \Lambda^{\bf A}&=\sum_{m_1,m_2} e^{-\pi t \left[
               m_1^2+{4\over 3}\left({m_1\over 2}-m_2\right)^2 \right]}, \cr
                \Lambda^{\bf B}&=\sum_{m_1,m_2} e^{-\pi t \left[
               m_1^2+12\left({m_1\over 2}-m_2\right)^2 \right]}. \cr }}
Finally, $n^{(ab)}_{\Sigma_1,\Sigma_2}$ denotes the trace of the action of
$\Sigma_2$ on  the fixed points in the $\Sigma_1$ twisted sector.
Taking into account that the origin is the only common fixed point
of $\ZZ_3$ and $\widehat{\ZZ}_3$, they can be determined to be
\eqn\anzahl{ n^{(ab)}_{\hat\Theta,\Omega}=\cases{
                        9& for $({\bf AA})$ \cr
                        3& for $({\bf AB})$ \cr
                        1& for $({\bf BB})$ \cr } }
and 
\eqn\anzahlb{ n^{(ab)}_{\hat\Theta,\Omega\Theta}=
              n^{(ab)}_{\hat\Theta^2,\Omega\Theta^2}=\cases{
                        -3& for $({\bf AA})$ \cr
                        i\sqrt{3}& for $({\bf AB})$ \cr
                        1& for $({\bf BB})$. \cr } }
The remaining numbers are given by complex conjugation of \anzahlb.
Applying a modular transformation to \klein\ yields the tree channel 
Klein bottle amplitude
\eqn\kleintree{\eqalign{ \tilde{K}^{(ab)}={32c\over 3} 
       \int_0^\infty {dl} 
      \biggl[ &n^a_0 n^b_0\, \rho_{00}\, \tilde\Lambda^a \tilde\Lambda^b + 
            {1\over 3} n^{(ab)}_{\hat\Theta^2,\Omega}\, \rho_{01} + 
            {1\over 3} n^{(ab)}_{\hat\Theta,\Omega}\, \rho_{02} +\cr
         & 3 \rho_{10} - 
             n^{(ab)}_{\hat\Theta^2,\Omega\Theta^2}\, \o\epsilon\rho_{11}-
             n^{(ab)}_{\hat\Theta,\Omega\Theta^2}\, \epsilon\rho_{12}+\cr
         & 3 \rho_{20} - 
             n^{(ab)}_{\hat\Theta^2,\Omega\Theta}\, \epsilon\rho_{21}-
             n^{(ab)}_{\hat\Theta,\Omega\Theta}\, \o\epsilon\rho_{22} \biggr].
          }} 
with $n_0^{\bf A}=\sqrt{3}$ and $n_0^{\bf B}=1/\sqrt{3}$. 
The lattice contributions are
 \eqn\zerosb{\eqalign{ \tilde\Lambda^{\bf A}&=\sum_{m_1,m_2} e^{-3\pi l \left[
               m_1^2+{4\over 3}\left({m_1\over 2}-m_2\right)^2 \right]}, \cr
                \tilde\Lambda^{\bf B}&=\sum_{m_1,m_2} e^{-\pi l \left[
               {1\over 3}m_1^2+4\left({m_1\over 2}-m_2\right)^2 \right]}. \cr }}
In order to cancel these tadpoles we now introduce D-branes with
mixed boundary conditions. For both the ${\bf A}$ and the ${\bf B}$
lattice we choose three kinds of D-branes with 
$\theta\in\{\pi/2,\pi/6,-\pi/6\}$. The asymmetric $\widehat{\ZZ}_3$ cyclically
permutes these three branes, whereas the symmetric ${\ZZ}_3$ leaves
every brane invariant and acts with a $\gamma_{\Theta,i}$ matrix on  the
Chan-Paton factors on each brane. Since $\widehat{\ZZ}_3$ permutes
the branes, all three $\gamma_{\Theta,i}$ actions must be the same.
The computation of the annulus amplitude gives 
\eqn\ann{\eqalign{ A^{(ab)}={c\over 12} \int_0^\infty {dt\over t^4} 
      \biggl[ &{\rm M^2}\rho_{00}\, \Lambda^a \Lambda^b + 
               ({\rm Tr}\, \gamma_\Theta )^2\,  \rho_{01} + 
               ({\rm Tr}\, \gamma_{\Theta^2} )^2\, \rho_{02} +\cr
            & {\rm M^2}\, n^{(ab)}_{\hat\Theta,1}\, \rho_{10}+
              ({\rm Tr}\, \gamma_\Theta )^2\,
               n^{(ab)}_{\hat\Theta,\Theta}\,\epsilon \rho_{11}+
             ({\rm Tr}\, \gamma_{\Theta^2} )^2\,
             n^{(ab)}_{\hat\Theta,\Theta^2}\,\o\epsilon \rho_{12}+\cr
            & {\rm M^2} \, n^{(ab)}_{\hat\Theta^2,1}\, \rho_{20}+
            ({\rm Tr}\, \gamma_\Theta )^2\,
            n^{(ab)}_{\hat\Theta^2,\Theta}\,\o\epsilon \rho_{21}+
           ({\rm Tr}\, \gamma_{\Theta^2} )^2\,
          n^{(ab)}_{\hat\Theta^2,\Theta^2}\,\epsilon \rho_{22} \biggr].}}
where the $\hat\theta$ twisted sector is given by open strings stretched
between D-branes with $\theta_i$ and $\theta_{i+1}$. 
Thus, $n^{(ab)}_{\hat\Theta,1}$ denotes the intersection number
of two such branes and $n^{(ab)}_{\hat\Theta,\Theta}$ the number
of intersection points invariant under $\Theta$.
The actual numbers turn out to be the same as the multiplicities of 
the closed string twisted sectors in \anzahl\ and \anzahlb.
For the tree channel amplitude we obtain
\eqn\anntree{\eqalign{ \tilde{A}^{(ab)}={c\over 6} 
       \int_0^\infty {dl} 
      \biggl[ &{\rm M}^2\, \left( n^a_0 n^b_0\, \rho_{00}\, 
           \tilde\Lambda^a \tilde\Lambda^b + 
            {1\over 3} n^{(ab)}_{\hat\Theta^2,1}\, \rho_{01} + 
            {1\over 3} n^{(ab)}_{\hat\Theta,1}\, \rho_{02}\right) +\cr
         & ({\rm Tr}\, \gamma_\Theta )^2 \left( 3 \rho_{10} - 
             n^{(ab)}_{\hat\Theta^2,\Theta}\, \o\epsilon\rho_{11}-
             n^{(ab)}_{\hat\Theta,\Theta}\, \epsilon\rho_{12}\right) +\cr
         & ({\rm Tr}\, \gamma_{\Theta^2} )^2 \left( 3 \rho_{20} - 
             n^{(ab)}_{\hat\Theta^2,\Theta^2}\, \epsilon\rho_{21}-
             n^{(ab)}_{\hat\Theta,\Theta^2}\, \o\epsilon\rho_{22}\right)
       \biggr].}}
Finally, one has to compute the M\"obius amplitude 
\eqn\moeb{\eqalign{ M^{(ab)}=-{c\over 12} \int_0^\infty &{dt\over t^4} 
      \biggl[ {\rm M}\, \rho_{00}\, \Lambda^a \Lambda^b + 
       {\rm Tr}( \gamma_{\Omega\Theta}^T\gamma_{\Omega\Theta}^{-1} )\, 
            \rho_{01} + 
             {\rm Tr}( \gamma_{\Omega\Theta^2}^T\gamma_{\Omega\Theta^2}^{-1} )
               \, \rho_{02} +\cr
            & {\rm M}\, n^{(ab)}_{\hat\Theta,\Omega}\, \rho_{11}+
              {\rm Tr}( \gamma_{\Omega\Theta}^T\gamma_{\Omega\Theta}^{-1} )\,
               n^{(ab)}_{\hat\Theta,\Omega\Theta}\,\epsilon \rho_{12}+
             {\rm Tr}( \gamma_{\Omega\Theta^2}^T\gamma_{\Omega\Theta^2}^{-1} )
          \,   n^{(ab)}_{\hat\Theta,\Omega\Theta^2}\,\o\epsilon \rho_{10}+\cr
     & {\rm M}\, n^{(ab)}_{\hat\Theta^2,\Omega}\, \rho_{22}+
              {\rm Tr}( \gamma_{\Omega\Theta}^T\gamma_{\Omega\Theta}^{-1} )\,
               n^{(ab)}_{\hat\Theta^2,\Omega\Theta}\,\o\epsilon \rho_{20}+
             {\rm Tr}( \gamma_{\Omega\Theta^2}^T\gamma_{\Omega\Theta^2}^{-1} )
      \,   n^{(ab)}_{\hat\Theta^2,\Omega\Theta^2}\,\epsilon \rho_{21} 
          \biggr],}}
with argument $q=-{\rm exp}(-2\pi t)$.
Transformation into tree channel leads to the expression
\eqn\moebtree{\eqalign{ \tilde{M}^{(ab)}=-{8c\over 3} 
       \int_0^\infty {dl} 
      \biggl[ &{\rm M}\, \left( n^a_0 n^b_0\, \rho_{00}\, 
           \tilde\Lambda^a \tilde\Lambda^b + 
            {1\over 3} n^{(ab)}_{\hat\Theta,\Omega}\, \rho_{01} + 
            {1\over 3} n^{(ab)}_{\hat\Theta^2,\Omega}\, \rho_{02}\right) +\cr
         &  {\rm Tr}( \gamma_{\Omega\Theta^2}^T\gamma_{\Omega\Theta^2}^{-1} )
           \left( 3 \rho_{11} - 
          n^{(ab)}_{\hat\Theta,\Omega\Theta^2}\, \o\epsilon\rho_{12}-
        n^{(ab)}_{\hat\Theta^2,\Omega\Theta^2}\, \epsilon\rho_{10}\right) +\cr
         &  {\rm Tr}( \gamma_{\Omega\Theta}^T\gamma_{\Omega\Theta}^{-1} )
            \left( 3 \rho_{22} - 
             n^{(ab)}_{\hat\Theta,\Omega\Theta}\, \epsilon\rho_{20}-
             n^{(ab)}_{\hat\Theta^2,\Omega\Theta}\, \o\epsilon\rho_{21}\right)
       \biggr].}} 
The three tree channel amplitudes give rise to two independent
tadpole cancellation conditions
\eqn\tadpole{\eqalign{  &{\rm M}^2-16\, {\rm M}+64=0, \cr
                        &({\rm Tr}\, \gamma_\Theta )^2-16 
            {\rm Tr}( \gamma_{\Omega\Theta^2}^T\gamma_{\Omega\Theta^2}^{-1} )
                 +64=0.}}
Thus, we have ${\rm M}=8$ D9-branes of each kind and the action
of $\ZZ_3$ on the Chan-Paton labels has to satisfy ${\rm Tr} \gamma_\Theta=8$
implying that we have the simple solution that $\gamma_\Theta$ is 
the identity matrix. 

\subsec{The massless spectrum}
\noindent
Having solved the tadpole cancellation conditions we can move forward
and compute the massless spectrum of the effective commutative field theory 
in the non-compact space-time.
In computing the massless spectra we have to take into account
the actions of the operations on the various fixed points. 
In  the closed string sector  we find the spectra shown in 
table 1
\vskip 0.2cm
\centerline{\vbox{
\hbox{\vbox{\offinterlineskip
\def\tablespace{height2pt&\omit&&\omit&&
 \omit&\cr}
\def\tablerule{\tablespace\noalign{\hrule}\tablespace}

\hrule\halign{&\vrule#&\strut\hskip0.2cm\hfil#\hfill\hskip0.2cm\cr
\tablespace
& $\epsilon$ && (ab) && spectrum       &\cr
\tablerule
& $1$ && $-$ && $(1,1)\,  {\rm Sugra}+4\times V_{1,1} $ &\cr
\tablerule
& $e^{\pm 2\pi i/3}$ && ${\bf AA}$ && $(0,1)\, {\rm Sugra}
+6\times T +15\times H$ &\cr
\tablespace
&  && ${\bf AB}$ && $(0,1)\, {\rm Sugra}+9\times T +12\times H$ &\cr
\tablespace
&  && ${\bf BB}$ && $(0,1)\, {\rm Sugra}+10\times T +11\times H$ &\cr
\tablespace}\hrule}}}}
\centerline{
\hbox{{\bf Table 1:}{\it ~~ closed string spectra}}}
\vskip 0.5cm
\noindent
The computation of the massless spectra in the open string sector
is also straightforward and yields the result in table 2
\vskip 0.2cm
\centerline{\vbox{
\hbox{\vbox{\offinterlineskip
\def\tablespace{height2pt&\omit&&\omit&&
 \omit&\cr}
\def\tablerule{\tablespace\noalign{\hrule}\tablespace}

\hrule\halign{&\vrule#&\strut\hskip0.2cm\hfil#\hfill\hskip0.2cm\cr
\tablespace
& $\epsilon$ && (ab) && spectrum       &\cr
\tablerule
& $1$ && $-$ && $V_{1,1}$ in $SO(8)$ &\cr
\tablerule
& $e^{\pm 2\pi i/3}$ && ${\bf AA}$ && $V$ in $SO(8)$ + $4\times H$ in
         ${\bf 28}$ &\cr
\tablespace
&  && ${\bf AB}$ && $V$ in $SO(8)$ + $1\times H$ in
         ${\bf 28}$ &\cr
\tablespace
&  && ${\bf BB}$ && $V$ in $SO(8)$  &\cr
\tablespace}\hrule}}}}
\centerline{
\hbox{{\bf Table 2:}{\it ~~ open string spectra}}}
\vskip 0.5cm
\noindent
All the spectra shown in table 1 and table 2 satisfy the cancellation
of the non-factorizable anomaly. Note, that the configurations ${\bf AB}$ 
and ${\bf BB}$
were not analyzed in \rbimopra. Thus, we have successfully applied
the techniques derived in section 2 and section 3 to the construction
of asymmetric orientifolds. 

\newsec{Conclusions}

In this article  we have pointed out a relationship between the
realization of asymmetric operations in the open string sector and 
noncommutative geometry arising at the boundary of the world-sheet
of open strings. More concretely, we have shown that  a left-right 
asymmetric rotation transforms an ordinary Neumann or 
Dirichlet boundary condition into a mixed Neumann-Dirichlet boundary condition. 
We have employed this observation to rederive the noncommutativity
relations for the open string. Moreover, we have solved the problem
of how the open string sector manages to incorporate asymmetric
symmetries. It simply turns on background gauge fluxes.
In asymmetric orbifolds different values of background gauge fields 
on the D-branes get identified and correspondingly different geometries, 
commutative or noncommutative, as well. 
Finally, we have considered a concrete asymmetric type I vacuum,
where D-branes with mixed boundary conditions were introduced
to cancel all tadpoles. 

We have restricted ourselves to the case of products of two-dimensional
tori. It would be interesting to generalize these ideas to more
general asymmetric elements of the T-duality group and to discuss the dual
heterotic description. 
Furthermore, it would be interesting to see whether via the asymmetric
rotation one can gain further insight into the relation between  
the effective noncommutative and commutative gauge theories 
on the branes.

\vskip 1cm

\centerline{{\bf Acknowledgements}}\pano
We would like to thank Volker Braun  and  Andreas Recknagel 
for encouraging discussions. The work is partially supported by the EU-TMR
project ERBFMRX-CT96-0045 and B.K. likes to thank the Studienstiftung 
des deutschen Volkes, as well. 
\bigno

\listrefs

\bye